\documentclass[preprint,superscriptaddress,12pt]{elsarticle}

\usepackage{amssymb}

\usepackage{amsmath}
\usepackage{multirow}
\usepackage{tabularx}
\usepackage{booktabs}
\usepackage{adjustbox}
\usepackage{threeparttable}
\usepackage{color}
\usepackage{float}

\usepackage{hyperref}

\journal{Science Bulletin}

\title{Evidence for particle acceleration approaching PeV energies in the W51 complex}

\begin{document}

\begin{frontmatter}
\cortext[cor1]{Corresponding authors\\}
\author{LHAASO Collaboration\corref{cor1}\fnref{fn1}}
\ead{lizhe@ihep.ac.cn(Zhe Li), fangkun@ihep.ac.cn(K.Fang), liuc@ihep.ac.cn,(C.Liu), silvia.celli@uniroma1.it (S. Celli)} 
\fntext[fn1]{Authors are listed at the end of this paper.}

\begin{abstract}
The $\gamma$-ray emission from the W51 complex is widely acknowledged to be attributed to the interaction between the cosmic rays (CRs) accelerated by the shock of supernova remnant (SNR) W51C and the dense molecular clouds in the adjacent star-forming region, W51B. However, the maximum acceleration capability of W51C for CRs remains elusive. Based on observations conducted with the Large High Altitude Air Shower Observatory (LHAASO), we report a significant detection of $\gamma$ rays emanating from the W51 complex, with energies from 2 TeV to 200 TeV. The LHAASO measurements, for the first time, extend the $\gamma$-ray emission from the W51 complex beyond 100 TeV and reveal a significant spectrum bending at tens of TeV. By combining the ``$\pi^0$-decay bump" featured data from Fermi-LAT, the broadband $\gamma$-ray spectrum of the W51 region can be well-characterized by a simple pp-collision model. The observed spectral bending feature suggests an exponential cutoff at $\sim400$~TeV or a power-law break at $\sim200$~TeV in the CR proton spectrum, most likely providing the first evidence of SNRs serving as CR accelerators approaching the PeV regime. Additionally, two young star clusters within W51B could also be theoretically viable to produce the most energetic $\gamma$ rays observed by LHAASO. Our findings strongly support the presence of extreme CR accelerators within the W51 complex and provide new insights into the origin of Galactic CRs.

\end{abstract}




\end{frontmatter}


\section{Introduction}
\label{sec:intro}
The W51 giant cloud stands out as one of the most massive and active star-forming regions in the Galaxy \cite{2017arXiv170206627G}. This cloud boasts an impressive diameter of $\sim 100$~pc and mass of $10^6 \, M_\odot$ \cite{1998AJ....116.1856C}. Within this expansive complex lie two primary clumps, W51A in the north and W51B to the west. Also located in this region, at a distance of $d=5.4-5.7$~kpc \cite{2018AJ....155..204R,2018ApJS..238...35S}, is the middle-aged supernova remnant (SNR) W51C (G$49.2-0.7$), most likely remnant of a very energetic core-collapse supernova explosion \cite{2005ApJ...633..946K}. W51C appears in radio as a shell-type SNR, with an estimated age of $t\sim30$~kyr \cite{1995ApJ...447..211K}, a kinetic energy release as high as $3.6 \times 10^{51}$~erg \cite{1995ApJ...447..211K}, and a radius of $24$~pc if $d=5.5$~kpc is assumed\footnote{There are still debates on the distance and age measurement of W51C. Some studies indicate that W51C is possibly located at a closer distance of 4.3 kpc \cite{Tian:2013zaa} and at a younger age of 18 kyr \cite{2017ApJ...849..147Z}.} \cite{1994JKAS...27...81M}. 

The detection of $\gamma$-ray radiation emanating from the W51 complex has been successively reported by a series of experiments, including H.E.S.S. \cite{Feinstein:2009zz}, Milagro \cite{Abdo:2009ku}, Fermi-LAT \cite{Fermi-LAT:2009qzy}, MAGIC \cite{MAGIC:2012anb}, and HAWC \cite{Abeysekara:2017hyn}. These observations span an energy range of nearly six orders of magnitude, from $\sim50$~MeV up to $\sim30$~TeV. The prevailing hypothesis for the source of these $\gamma$ rays is the interaction between cosmic-ray (CR) nuclei accelerated at the W51C shock and the nearby molecular clouds (MCs) in W51B, through the neutral pion ($\pi^0$) decay channel. Support for this scenario is multifaceted, including extensive evidence of SNR-MC interactions observable in radio bands \cite{1997ApJ...475..194K,1997ApJ...485..263K,1997AJ....114.2058G,2014ApJ...786L..24D} and the identification of the spectral feature ``$\pi^0$-decay bump" \cite{Jogler:2016lav}. The latter is considered a distinctive signature of CR-MC interactions, representing a crucial piece of evidence for probing CR acceleration in SNR-driven shocks. In $\log-\log$ coordinates, the $\gamma$-ray energy spectrum exhibits a symmetry relative to half the rest mass of $\pi^0$, thereby creating such a sub-GeV bump structure. W51C is one of the few SNRs with a clear $\pi^0$-decay bump detected \cite{Fermi-LAT:2013iui,Jogler:2016lav,HESS:2016qan}.

SNRs have long been considered the major factories of Galactic CRs \cite{1961PThPS..20....1G}, owing to the fact that supernova explosions can provide sufficient energy to maintain the observed CR intensity, with a required conversion efficiency of $\approx10\%$, and to the presence of shock regions as reliable sites for accelerating CRs \cite{1981ICRC...12..155A}. However, because $\gamma$-ray spectral measurements of SNRs have been constrained to energies well below $100$ TeV until recently, there has been a lack of definitive evidence to establish SNRs as accelerators CRs with energies of PeV, where the \emph{knee} of the CR energy spectrum is located \cite{LHAASO:2024knt}, a feature that may represent the upper limit of acceleration by Galactic sources.
Therefore, the detection of ultra-high-energy (UHE, $>100$~TeV) $\gamma$ rays from SNRs, particularly from those like W51C with established CR acceleration evidence, presents an excellent opportunity to study the acceleration potential of SNRs and the origin of Galactic CRs \cite{Chen:2022eqp}.


On the other hand, the intricate environment of W51 harbors additional potential sources of UHE $\gamma$ rays. Among these is the pulsar wind nebula (PWN) candidate, CXO J192318.5$+$140305 \cite{2005ApJ...633..946K}, which has possibly originated from the same explosion that produced W51C. Its hard X-ray emission \cite{2005ApJ...633..946K,Hanabata:2012qb,Sasaki:2014gda} suggests that it has the potential to accelerate high-energy leptons. These leptons, in turn, could be responsible for generating TeV $\gamma$ rays through the inverse Compton (IC) scattering process. Furthermore, the W51 complex hosts several young star clusters (YSCs) \cite{2004MNRAS.353.1025K}. They have been suggested as alternative contributors to SNRs recently \cite{2019NatAs...3..561A}, which is supported by the detection of several PeV photons in the Cygnus region by LHAASO \cite{LHAASO:2023pum}.

In this work, we report the Large High Altitude Air Shower Observatory (LHAASO) measurements of the $\gamma$-ray spectrum of the W51 complex, covering an impressive energy range from $2$~TeV to the UHE domain of $200$~TeV. We first detail the LHAASO observations and the corresponding analytical outcomes. Subsequently, we deliberate on the discussion of physical scenarios that aim to interpret the broadband $\gamma$-ray spectrum of this astrophysically rich region.

\section{Observation}
\label{sec:result}
\subsection{Experiment}
\label{subsec:obser}

LHAASO is a ground-based experiment located high on the edge of the Tibetan Plateau at an average altitude of 4410 meters. It has broken through the limited energy range of space-borne instruments and sensitively extended the measured energy beyond sub-TeV to PeV \cite{deNaurois:2015oda}. This hybrid array consists of the Kilometer Square Array (KM2A), the Water Cherenkov Detector Array (WCDA), and the Wide Field of View Cherenkov Telescope Array (WFCTA)~\cite{LHAASO:2019qwt}. With a large field of view and continuous exposure, WCDA and KM2A provide full-duty monitoring of $\gamma$-ray emissions from almost $60\%$ of the sky each day \cite{Aharonian:2020iou}. The $\gamma$/hadron separation capability of KM2A in the UHE band is better than $4\times 10^3$ to reject the contamination of CR-induced showers \cite{Aharonian:2020iou}. Performance study of KM2A indicates the energy resolution is better than 20\% and angular resolution is smaller than $0.25^\circ$ at 100~TeV \cite{Aharonian:2020iou}. All these features allow KM2A to achieve $1\%$ sensitivity above 100 TeV for Crab-like sources in one year \cite{lhaaso:whitepaper1},  and to precisely determine the $\gamma$-ray spectral characteristics of SNRs with unprecedented sensitivity beyond 100 TeV \cite{Chen:2022eqp}. 

We conducted measurements with LHAASO, including both WCDA and KM2A in the W51 region over effective exposure times of 795.96 days (from March 5, 2021, to July 31, 2023) and 1216.24 days (from December 27, 2019, to July 31, 2023), respectively. Analogous data-processing criteria were adopted as previous works \cite{Aharonian:2020iou,LHAASO:2023rpg}, including the detector trigger model, noise filter method, event direction selection, $\gamma$/hadron separation efficiency, and detector response simulation. For background estimation, we employed the Direct Integration method \cite{Atkins_2003}, which has been widely used in $\gamma$-ray source surveying with LHAASO \cite{Aharonian:2020iou,LHAASO:2021gok}.

\subsection{Morphology}
\label{subsec:detection}

The spatial and spectral distribution of radiation from the W51 region was estimated by employing a conservative approach by fitting a relatively large region of interest (ROI), with declination (Dec.) from $11^\circ$ to $19^\circ$ and Right ascension (R.A.) from $286.5^\circ$ to $294.9^\circ$, covering the entire complex and the ambient diffuse regions. We modeled the contamination due to the Galactic diffuse emission (GDE) consistently with the spatial distribution of Planck dust observations \cite{Planck:2013ltf,Planck:2016frx}, applying the same flux treatment as established in previous works \cite{LHAASO:2023rpg}. We further considered two significant spatially extended nearby sources, LHAASO~J1919$+$1556 ($1\sigma$ extent $\sigma_{\rm ext}=0.11^{\circ}\pm0.04^{\circ}_{\rm stat}$) and LHAASO~J1924$+$1609 ($1\sigma$ extent $\sigma_{\rm ext}=1.47^{\circ}\pm0.06^{\circ}_{\rm stat}$), and subtracted their contributions. Other sources located at larger distances from the W51 region were subtracted using the same source parameters as provided in the LHAASO catalog \cite{LHAASO:2023pum}.

\begin{figure}[H]
    \centering
    \begin{adjustbox}{width=1.2\textwidth,center}
    \begin{minipage}{0.6\textwidth}
    \includegraphics[width=\linewidth, trim={1cm 0cm 1cm 2cm}, clip]{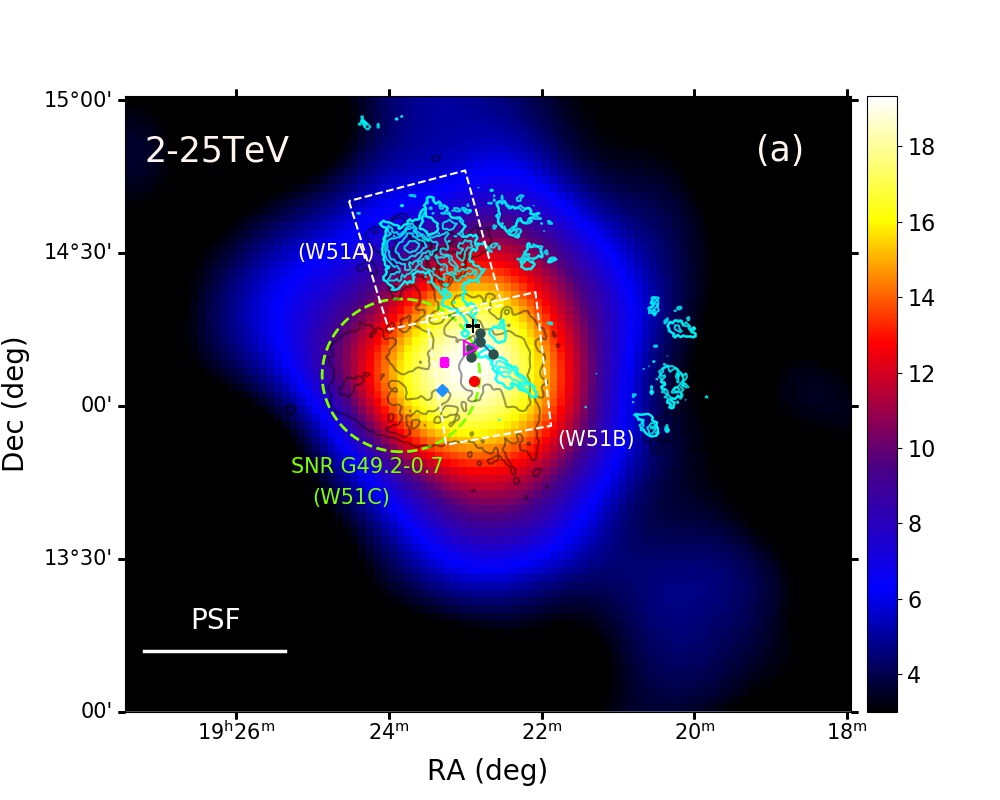}
    \end{minipage}%
    \begin{minipage}{0.6\textwidth}
    \includegraphics[width=\linewidth, trim={1cm 0cm 1cm 2cm}, clip]{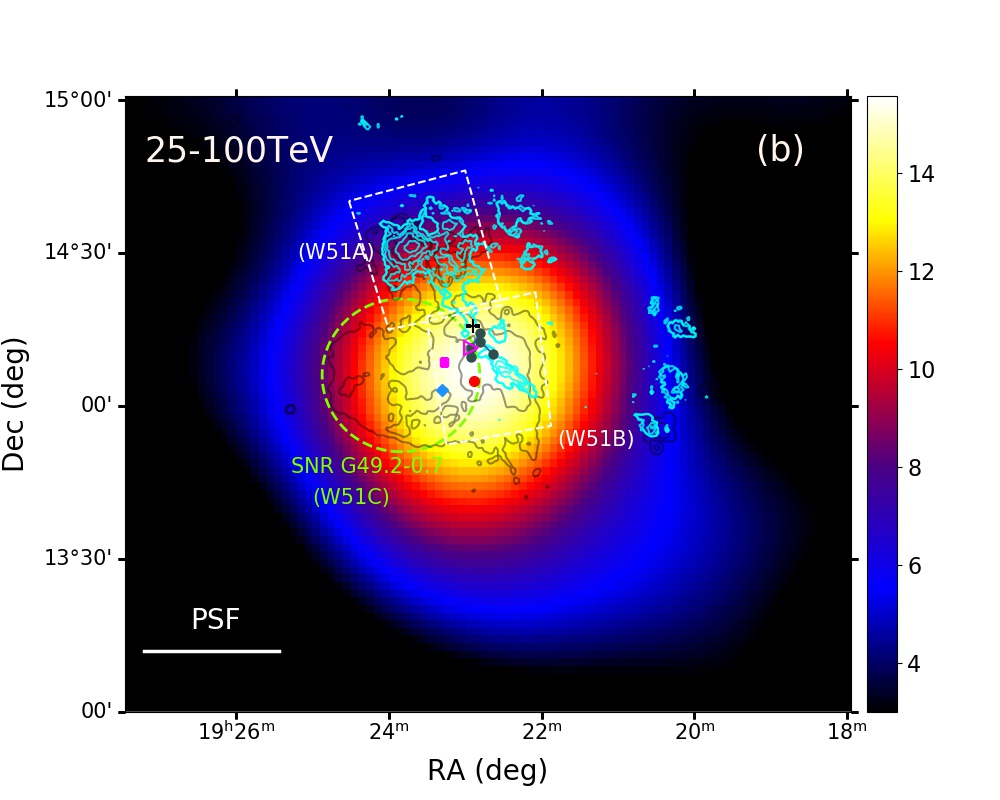}
    \end{minipage}
    \end{adjustbox}

    \begin{adjustbox}{width=0.6\textwidth, center}
    \begin{minipage}{\textwidth}
    \centering
    \includegraphics[width=\linewidth, trim={1cm 0cm 1cm 2cm}, clip]{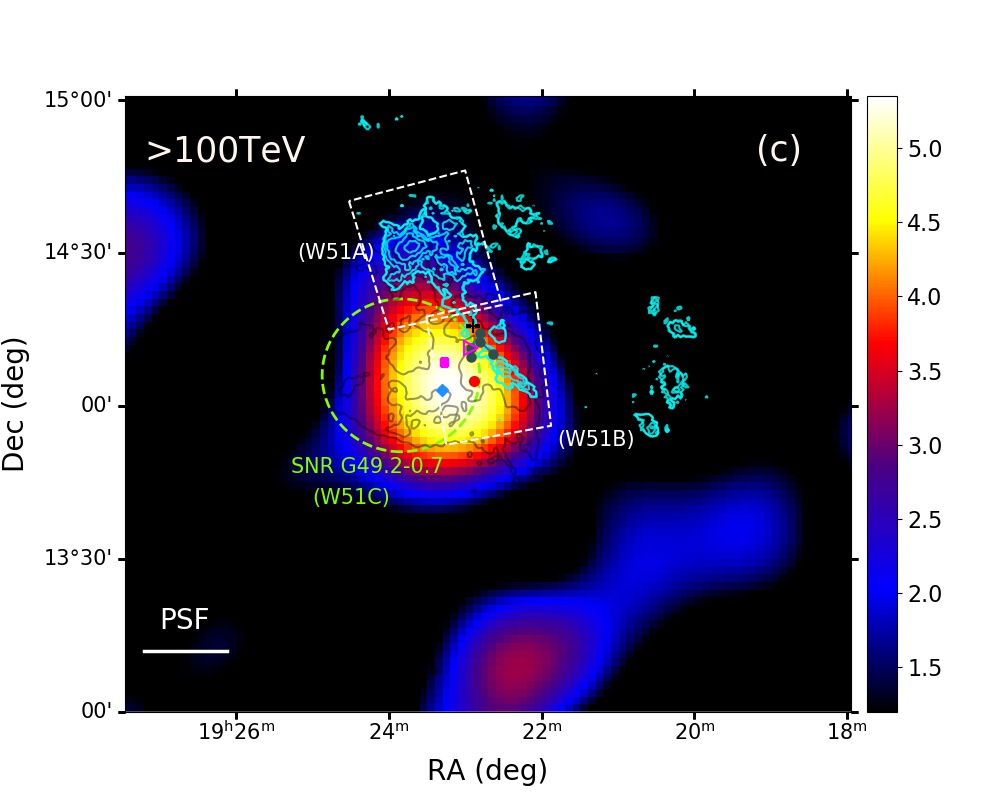}
    \end{minipage}
    \end{adjustbox}
    \caption{Gamma-ray maps in the intervals $[2-25]$(a), $[25-100]$~TeV(b) and $>100$~TeV(c) of the W51 region presented in equatorial coordinates. The color scale indicates the statistical significance of the excess $\gamma$-ray counts after subtracting the Galactic diffuse emission. Maps have been smoothed with a Gaussian kernel of $1.58$ times the PSF. The central position determined by the all-energy-range fit to the LHAASO data is marked by a red dot. The $\gamma$-ray centroids given by Fermi-LAT \cite{Fermi-LAT:2009qzy} and MAGIC \cite{MAGIC:2012anb} observations are marked with a magenta square and a magenta triangle, respectively. Cyan contours show the MWISP \cite{WMISP} measurement of the ${}^{12}$CO emission integrated the velocity from 54 to 70 km s$^{-1}$. Black contours overlay the 1.4~GHz continuum emission as observed by the Very Large Array \cite{Stil_2006}. The blue diamond marks the position of a PWN candidate CXO J192318.5$+$140305 \cite{2005ApJ...633..946K}. The black cross shows where 1720 MHz OH masers are emitted from \cite{Brogan:2013jba}, while six dark-green marked points localize shocked CO clumps \cite{1997ApJ...475..194K,1997ApJ...485..263K}. Two white dashed rectangles define the two star-forming regions, W51A and W51B \cite{fujita2019massive}. The radio shell of SNR W51C is indicated with a green dashed circle \cite{Brogan:2013jba}.(Color online)}
    \label{fig:skymap}
\end{figure}

The spatial distribution and energy spectrum of the W51 complex are simultaneously measured by means of a binned maximum 3D-likelihood analysis method \cite{LHAASO:2023rpg}. Gamma-ray excesses with statistical significances of $19.33\sigma$, $15.58\sigma$, and $5.35\sigma$ are revealed across three energy regimes of $<25$~TeV, $25-100$~TeV, and $>100$~TeV, respectively. Gamma-ray significance maps of these three energy bins are illustrated in Fig.~\ref{fig:skymap}. By adopting a 2D Gaussian distribution template for the $\gamma$-ray emission, the overall centroid was found at $({\rm R.A.,Dec.})=(290.72^\circ\pm0.02^\circ_{\rm stat},14.08^\circ\pm0.02^\circ_{\rm stat})$, which is shown in Fig.~\ref{fig:skymap} with a red dot. The overall $\gamma$-ray emission exhibits an extension with $\sigma_{\rm ext} =0.17^\circ\pm0.02^{\circ}_{\rm stat}$ after removing the Point Spread Function (PSF), consistent with previous findings from Fermi-LAT and MAGIC considering statistical uncertainties \cite{Fermi-LAT:2009qzy,MAGIC:2012anb}.  

The radio continuum map at 1.4 GHz, which delineates the components of the W51 complex, is also illustrated in Fig.~\ref{fig:skymap} \cite{Stil_2006}. It is evident that the $\gamma$-ray emission region observed by LHAASO coincides with W51C and W51B, whereas no significant emission is detected from W51A. There are two compelling pieces of evidence for the interaction between the shock driven by W51C and the molecular cloud associated with W51B.
One is the detection of OH masers at 1720~MHz \cite{1997AJ....114.2058G,Brogan:2013jba}, and the other is the detection of shocked CO clumps and atomic gas in the adjacent region between W51C and W51B \cite{1997ApJ...475..194K,1997ApJ...485..263K},  both of which are spatially consistent with the centroid of the $\gamma$-ray emission measured by LHAASO. In addition, we superimposed the high-resolution $^{12}$CO contour provided by the Milky Way Imaging Scroll Painting (MWISP) survey \cite{WMISP} onto the $\gamma$-ray map to provide a more detailed context of the MC environment.



According to LHAASO observations and analysis, the PSF sizes at energy intervals of $2-25$~TeV, $25-100$~TeV, and above $100$~TeV are $0.46^\circ$, $0.44^\circ$, and $0.27^\circ$, respectively. Given that these values significantly exceed the angular separation between the SNR-MC interaction region and the putative PWN mentioned above, morphological analysis alone is insufficient to distinctly attribute $\gamma$-ray emission to these sources independently. Nonetheless, a detailed physical assessment of the potential contribution of the PWN candidate to the UHE $\gamma$-ray emission is provided in Appendix B.



\subsection{Spectrum}
\label{subsec:spectr}
In the estimation of the $\gamma$-ray spectrum, our initial 3D-likelihood fitting approach consists of adopting a simulated detector response to convert the observed number of excess events ($N_s$) into differential flux points \cite{LHAASO:2023rpg}, based on an assumed spectrum model such as pure power-law function, power-law with exponential cutoff (PLExpCut), or log-parabola (LOG). Energy resolution and the energy bias of both WCDA and KM2A thus have to be taken into account by unfolding the spectrum \cite{Aharonian:2020iou}. 

\begin{figure}[h!]
    \centering
    \includegraphics[width=0.8\textwidth]{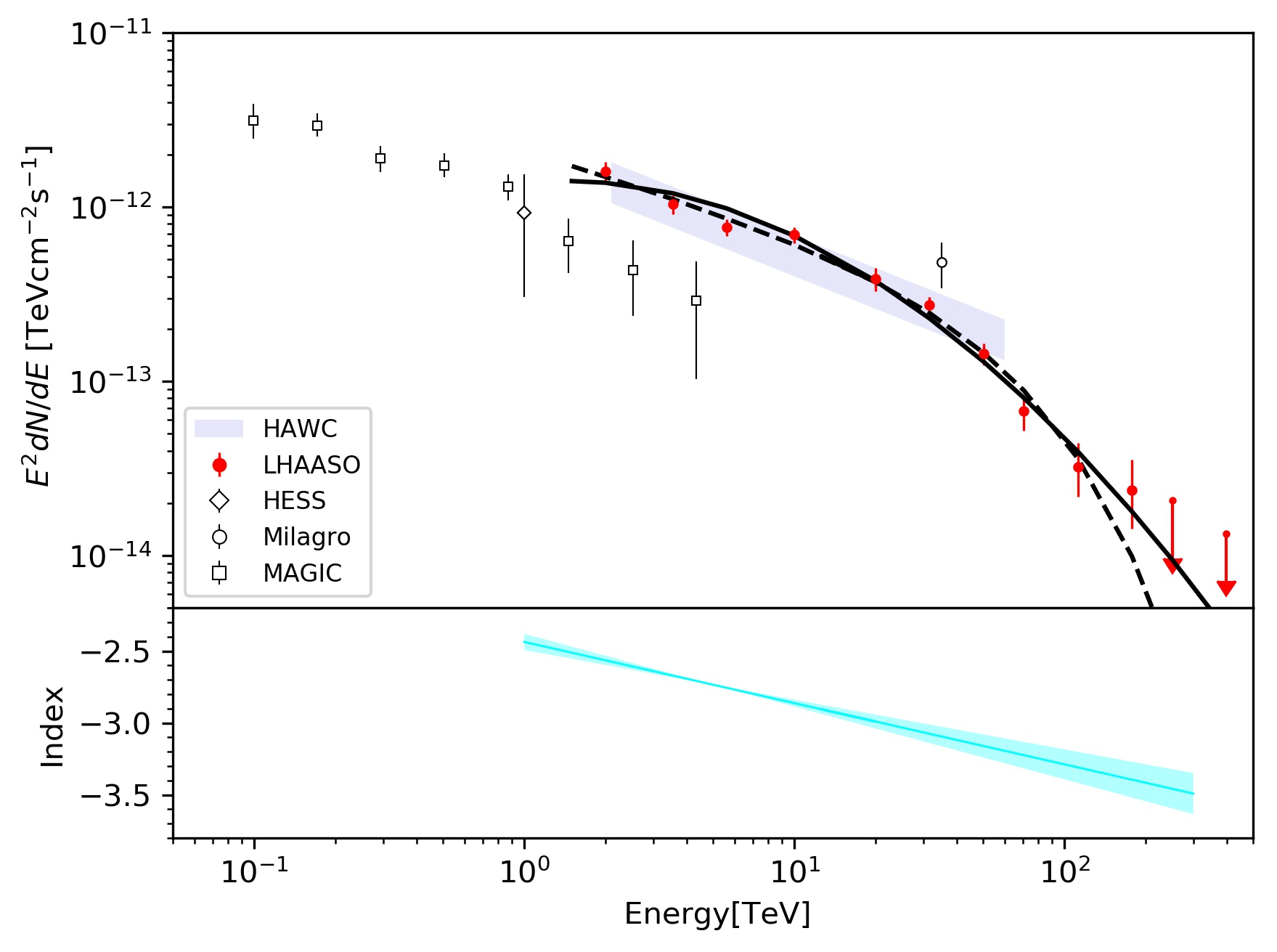}
    \caption{Spectral energy distribution of W51C plotted as $E_\gamma^2dN_\gamma/dE_\gamma$. \emph{Top panel:} Red flux points, covered an energy band from $2.24^{+3.01}_{-1.28}$~TeV to $177.83^{+46.04}_{-36.57}$~TeV, are the measured fluxes from LHAASO by performing a 3D-likelihood method assuming a PLExpCut spectrum in the simulation. The error bar for all points are given at 1 times standard deviation. The statistical significance at the highest detection point is 3.33$\sigma$. The last two flux upper limits are given at $\simeq281.84^{+72.98}_{-57.97}$~TeV and $\simeq446.68^{+102.86}_{-83.61}$~TeV, respectively. Previous observation results from MAGIC \cite{MAGIC:2012anb}, H.E.S.S. \cite{HESS:2016qan}, Milagro \cite{Abdo:2009ku}, and HAWC \cite{HAWC:2020hrt} are also depicted. The solid black line is the fitting result with a LOG function, while the dashed black line represents the fitting result with a PLExpCut function. \emph{Bottom panel:} Energy-dependent $\gamma$-ray slope as derived by the LOG model. The cyan band represents the $1\sigma$ confidence interval.(Color online)}
    \label{fig:sedmap}
\end{figure}

The $\gamma$-ray spectrum of the W51 complex, as depicted in Fig.~\ref{fig:sedmap}, extends from $\approx2$~TeV to an unprecedented upper limit of $\approx200$~TeV, marking the first observation of this astrophysical object in the UHE band. The spectrum below $25$ TeV has been precisely measured by WCDA and is consistent with the result from HAWC \cite{HAWC:2020hrt}, another Extensive Air Shower (EAS) facility. Notably,the flux measured by LHAASO below 10~TeV is about 1.8 times higher than that reported by MAGIC \cite{MAGIC:2012anb}.
When the different spatial extensions observed by MAGIC and LHAASO are accounted for, the flux should exhibit an increase by a factor of 1.3 relative to the MAGIC data. Systematic biases between imaging telescopes and EAS arrays for point-like sources may also be a factor contributing to this difference, as indicated by the spectrum measurements of the Crab Nebula \cite{LHAASO:2021cbz}. Additionally, flux points from H.E.S.S. and Milagro also exhibit slight differences while remaining statistically consistent with the current measurements.

The $\gamma$-ray spectrum above tens of TeV is characterized by a significant curvature: as such, to quantify the spectral index variation as a function of photon energy, we employed the LOG function in our 3D likelihood estimation. The spectral model is expressed as follows:
\begin{equation}
\frac{dN_{\gamma}}{dE_\gamma}=J_0(E_{\gamma}/20\,{\rm TeV})^{-[a+b{\rm log_{10}}(E_{\gamma}/20\,{\rm TeV})]} \,.
\label{func:log}
\end{equation}
The estimated parameters are $J_0=(0.94\pm 0.06_{\rm stat})\times10^{-15}$~cm$^{-2}$~s$^{-1}$~TeV$^{-1}$, $a=2.98\pm0.04_{\rm stat}$, and $b=0.43\pm0.08_{\rm stat}$. The goodness of fit of the model to the corresponding flux points yields $\chi^2/ndf=10.43/7$. The confidence interval of $b$ indicates that the $\gamma$-ray spectrum deviates from a power-law function with a $5.4\sigma$ significance, exhibiting a very pronounced spectral bending.
It is important to note that the best-fit spectrum obtained from the 3D likelihood approach may not coincide precisely with the spectrum that would result from a direct fit to the flux points depicted in Fig.~\ref{fig:sedmap}. 

If the UHE emission is attributed to proton-proton collisions, the detection of photons with energies of a few hundred TeV may imply the responsible source to be a PeVatron \cite{LHAASO:2023pum}. The maximum acceleration limit for protons may manifest in the $\gamma$-ray spectrum as a characteristic exponential cut-off feature.
Thus, we also adopt a PLExpCut function to describe the $\gamma$-ray spectrum, which is expressed as
\begin{equation}
 \frac{dN_{\gamma}}{dE_\gamma}=J_0(E_{\gamma}/20\,{\rm TeV})^{-\Gamma} \exp(-E_{\gamma}/E_{\gamma,\rm cut})\,.
 \label{func:cut}
\end{equation}
The parameters estimated by the 3D-likelihood method are $J_0=(1.29\pm 0.18_{\rm stat})\times10^{-15}$~cm$^{-2}$~s$^{-1}$~TeV$^{-1}$, $\Gamma$=$2.48\pm0.08_{\rm stat}$, and $E_{\gamma,\rm cut}$=$ (61.05\pm 15.28_{\rm stat})$~TeV. The goodness of fit characterized by the chi-square statistic is $\chi^2/ndf=7.94/7$. The PLExpCut function marginally underestimates the flux around 200 TeV, suggesting that a model with a smoother cutoff may yield a better representation \cite{celli2020}. 

Owing to the unprecedented sensitivity of LHAASO at the UHE band, the bending of the $\gamma$-ray spectrum in the W51 complex is detected for the first time. The cutoff energy $E_{\gamma,\rm cut}$ derived from the PLExpCut fitting clearly indicates the spectral bending at tens of TeV, corresponding to the spectral softening revealed by the LOG model. The bending feature is a clear deviation from the pure power-law spectrum, suggesting that we have detected a signature of either the maximum acceleration or the confinement capability of the source in this region.

The systematic uncertainties in the $\gamma$-ray spectrum measurements of the LHAASO experiment have been thoroughly investigated in \cite{Aharonian:2020iou,LHAASO:2021WCDAcrab}, leading to $7\%$ and $^{+8\%}_{-24\%}$ for the KM2A and WCDA measurements, respectively. The primary uncertainty arises from the atmospheric model used in the Monte Carlo simulations. Furthermore, we have evaluated the uncertainties arising from the GDE and the influence of nearby sources, determining that the flux may vary by $13\%$ at maximum. Variations in the initial spectral models used for simulation, which in turn influence the generated detector response, can induce a fluctuation of about $1\%-9\%$ in the final flux estimation. The positional accuracy of our results is subject to a systematic pointing uncertainty of $\approx0.04^\circ$, as established by targeting point-like sources such as the Crab Nebula, Mrk 421, and Mrk 501 \cite{LHAASO:2023rpg}. Similarly, the systematic uncertainty of the source extension was estimated to be $\approx0.05^\circ$\cite{LHAASO:2023rpg}.

\subsection{Joint fitting of LHAASO and Fermi-LAT data with a simple hadronic model}
\label{subsec:hadronic}
\begin{figure}[h!]
  \centering
  \begin{minipage}{0.5\textwidth}
  \includegraphics[width=\linewidth, trim={0cm 0cm 1cm 1cm}, clip]{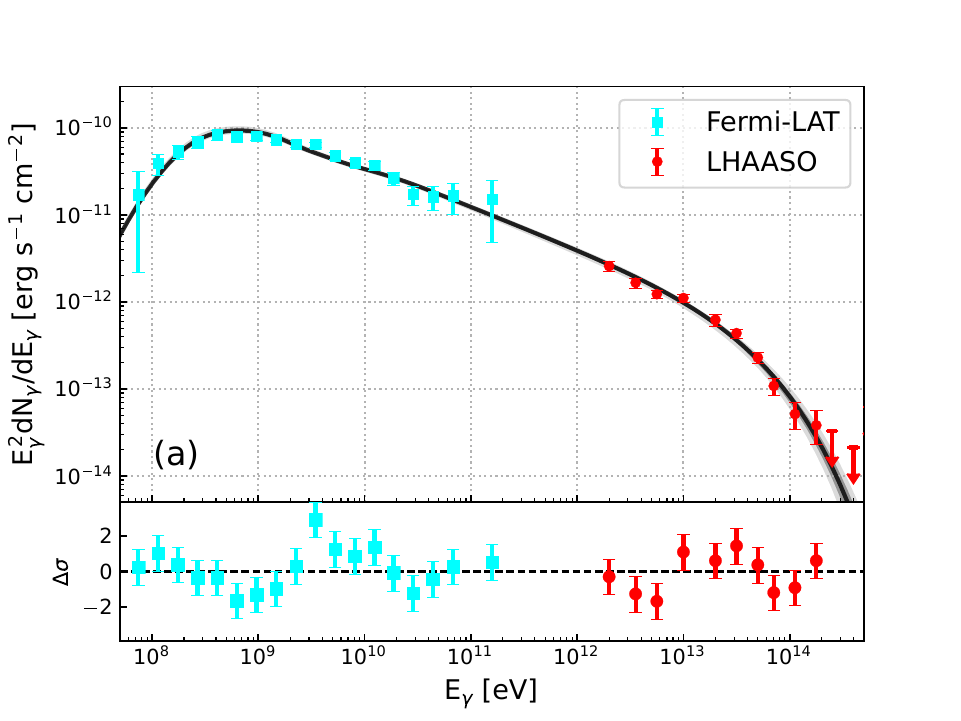}
  \end{minipage}%
  \begin{minipage}{0.5\textwidth}
  \includegraphics[width=\linewidth, trim={0cm 0cm 1cm 1cm}, clip]{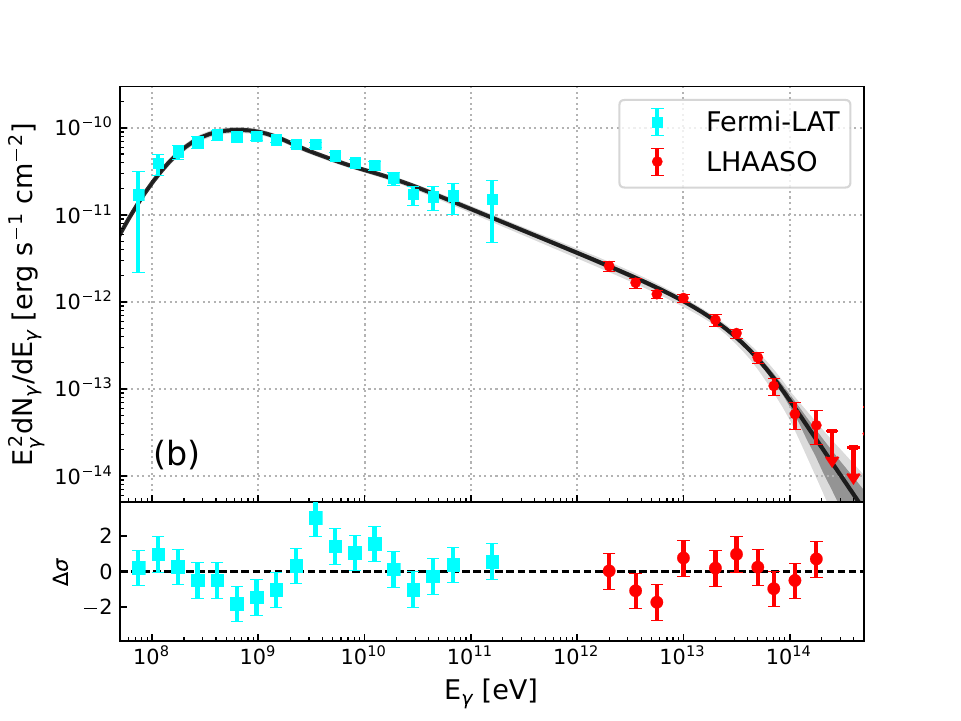}
  \end{minipage}
  \caption{Fitting results of the $\pi^0$-decay model to the $\gamma$-ray spectrum measurements of W51C. The LHAASO spectrum reported by the present work and the Fermi-LAT spectrum \cite{Jogler:2016lav} are used in the fitting processes. The proton incident spectrum takes the form of ECPL (BPL) in the left (right) panel. For each panel, the main plot presents the differential energy spectrum of $\gamma$ rays, denoted by ${\rm d}N_\gamma/{\rm d}E_\gamma$. The black line is the best-fit curve, while the dark and light grey bands are the $2\sigma$ and $3\sigma$ confidence intervals of the model, respectively. Directly beneath, a subplot details the standardized residuals between the best-fit curve and the data.(Color online)}
  \label{fig:model_pp}
\end{figure}

The presence of the $\pi^0$-decay bump unambiguously signifies a hadronic mechanism behind the $\gamma$-ray emission observed by Fermi-LAT \cite{Jogler:2016lav}. Considering that the $\gamma$-ray emission detected by LHAASO is spatially consistent with that reported by Fermi-LAT, and the energy spectrum measurements of the two instruments can be smoothly bridged with a power-law form (as will be illustrated in Fig.~\ref{fig:model_pp}), it is reasonable to ascribe the broadband $\gamma$-ray energy spectrum to a unified $\pi^0$-decay framework.

Computation is performed using the Python package Naima \cite{naima}. The $\gamma$-ray flux generated by $\pi^0$ decay is dependent on the proton incident flux, the target gas density, and the $\pi^0$-production cross-section. We adopt for the energy spectrum of the incident protons a power law function with an exponential cut-off (ECPL), described as
\begin{equation}
 \frac{dN_p}{dE_p}\propto E_p^{-\alpha}\cdot{\rm exp}\left(-\frac{E_p}{E_{p,\rm cut}}\right)~,
 \label{eq:proton_ecpl}
\end{equation}
where $E_p$ is the proton energy, and $dN_p/dE_p$ is the energy differential number density of protons. Alternatively, broken power law (BPL) could also be a plausible form for the proton spectrum, expressed as
\begin{equation}
 \frac{dN_p}{dE_p}\propto\left\{
 \begin{aligned}
  & E_p^{-\alpha_1}, & E_p<E_{p,\rm br} \\
  & E_{p,\rm br}^{\alpha_2-\alpha_1}E_p^{-\alpha_2}, & E_p>E_{p,\rm br} \\
 \end{aligned}
 \right.~.
 \label{eq:proton_bpl}
\end{equation}

The gas in the interaction region is predominantly atomic hydrogen that has been dissociated from molecular hydrogen \cite{1997ApJ...475..194K}. As the density of the expanding atomic hydrogen gas pushed by the SNR shock is estimated to be $\approx30-160$~cm$^{-3}$ (derived from the scale and column density of the expanding gas \cite{1997ApJ...475..194K}), we set the average of this range as the default value, which is $n_H\approx100$~cm$^{-3}$. The Pythia 8 option embedded in the Naima package \cite{Kafexhiu:2014cua} is used for the $\pi^0$-production cross section. The free parameters for the ECPL scenario are $\alpha$, $E_{p,\rm cut}$, and the total energy of the proton incident spectrum above $1$~GeV, denoted by $W_p$, while those for the BPL scenario are $\alpha_1$, $\alpha_2$, $E_{p,\rm br}$, and $W_p$. Markov Chain Monte Carlo fitting \cite{Foreman-Mackey:2012any} is applied for parameter estimation.

The fitting results are depicted in Fig.~\ref{fig:model_pp}. The reduced $\chi^2$ statistic of the best-fit model is $32.4/25$ for the scenario of the ECPL proton spectrum while $31.8/24$ for the BPL case. The difference in the Bayesian information criterion\footnote{The definition is $-2\ln L+k\ln n$, where $L$ is the likelihood, $k$ is the number of free parameters, and $n$ is the number of data points.} between the former and the latter is $\approx-3$, indicating that neither model demonstrates significant superiority over the other.

\begin{table*}[h!]
  \centering
  \caption{Parameter estimation results obtained by fitting LHAASO and Fermi-LAT data with simple hadronic models. The upper section shows the results for an ECPL incident proton spectrum, while the lower corresponds to the BPL case. For each parameter, the table lists both the maximum a posterior (MAP) estimate and the median of the posterior distribution (accompanied by the 16th and 84th percentile confidence interval). The reduced $\chi^2$ is calculated using the MAP parameters.}
  \scriptsize
  \begin{threeparttable}  
  \setlength{\extrarowheight}{3pt}  
  \begin{tabular}{cccccccccc}
    \toprule
    \multicolumn{2}{c}{$\alpha$} & & \multicolumn{2}{c}{$E_{\rm cut}$~[TeV]} & & \multicolumn{2}{c}{$W_{p,0}$\textsuperscript{\dag}  [$10^{50}$~erg]} & & $\chi^2/ndf$ \\
    \cline{1-2} \cline{4-5} \cline{7-8} \cline{10-10} 
    Best & Median & & Best & Median & & Best & Median & &   \\ 
    $2.55$ & $2.55_{-0.01}^{+0.01}$ & & $382$ & $383_{-54}^{+66}$ & & $1.30$ & $1.30_{-0.04}^{+0.04}$ & & $32.4/25$ \\ 
    \bottomrule
  \end{tabular}   
  \textsuperscript{\dag}$W_{p,0}$ is the total proton energy when the default values of $d$ and $n_H$ are adopted, that is, $W_p=W_{p,0}(d/5.5\,{\rm kpc})^2(n_H/100\,{\rm cm}^{-3})^{-1}$.
  \end{threeparttable}

  \begin{threeparttable}  
  \setlength{\extrarowheight}{3pt}  
  \begin{tabular}{ccccccccccccc}
    \toprule
    \multicolumn{2}{c}{$\alpha_1$} & & \multicolumn{2}{c}{$\alpha_2$} & & \multicolumn{2}{c}{$E_{\rm br}$~[TeV]} & & \multicolumn{2}{c}{$W_{p,0}$ [$10^{50}$~erg]} & & $\chi^2/ndf$ \\
    \cline{1-2} \cline{4-5} \cline{7-8} \cline{10-11} \cline{13-13} 
    Best & Median & & Best & Median & & Best & Median & & Best & Median & &   \\ 
    $2.56$ & $2.56_{-0.01}^{+0.01}$ & & $3.86$ & $4.07_{-0.44}^{+0.55}$ & & $161$ & $179_{-45}^{+42}$ & & $1.32$ & $1.33_{-0.04}^{+0.04}$ & & $31.8/24$ \\ 
    \bottomrule
  \end{tabular}   
  \end{threeparttable}
  
  \label{tab:params}
\end{table*}

The estimated parameters are $\alpha=2.55\pm0.01$ and $E_{p,\rm cut}=385_{-55}^{+65}$~TeV for the scenario of the ECPL proton spectrum, and $\alpha_1=2.56\pm0.01$, $\alpha_2=4.07_{-0.44}^{+0.55}$, and $E_{p,\rm br}=(180\pm45)$~TeV for the BPL proton spectrum. In both cases, the total energy of the proton spectrum is $W_p=(1.30\pm0.05)\times10^{50}~(d/5.5\,{\rm kpc})^2(n_H/100\,{\rm cm}^{-3})^{-1}$~erg. These results are summarized in Table~\ref{tab:params}. Remarkably, the values of $E_{p,\rm cut}$ and $E_{p,\rm br}$ indicate that a considerable number of CRs have been accelerated beyond $100$~TeV. With the default values of $d$ and $n_H$, the required total proton energy amounts to a mere 4\% of the kinetic energy of SNR W51C. This energy proportion is well within a feasible supply range for SNRs. Additionally, we have checked that the results of parameter estimation are slightly affected by the choice of different hadronic interaction models while remaining consistent within the range of error.



\section{Interpreting the origin of the radiation}
\label{sec:discuss}

The W51 complex is quite rich in candidate particle accelerators, possibly producing radiation extending in the UHE domain, among which are the SNR W51C, the putative PWN from the same explosion, and several YSCs. We present a detailed investigation of each scenario in the following. 

\subsection{The supernova remnant W51C}
The middle-aged SNR W51C is a well-known $\gamma$-ray emitter, belonging to one of the few cases where the hadronic origin of its low-energy radiation could be ascertained thanks to the clear detection of the $\pi^0$-bump. The hadronic scenario is supported by the presence of massive nearby MCs providing a dense enough target for hadronic collisions to occur, such that the resulting secondary $\gamma$-ray flux is enhanced. 
Indeed, strong and multiple pieces of evidence indicate the ongoing collision between the W51C shock and massive clouds within the surrounding giant MC. As such, the most likely origin of the observed $\gamma$-ray emission up to the highest energies can be realized in the scenario of the interaction between the shock-accelerated particles and the nuclei in the clouds, either in direct collisions \cite{Fang:2010rq} or after their nominal escape from the acceleration cycles \cite{1996A&A...309..917A,2007Ap&SS.309..365G,celli2019}.

In the direct interaction scenario, namely among freshly accelerated particles and target clouds, the slope of the incident proton spectrum inferred from the data would correspond to the acceleration stage: its value of $\alpha\approx2.5$ may correspond to a low Mach number shock ($M\approx3$), consistently with the expectation for a middle-aged SNR like W51C \cite{Fang:2010rq}. However, the observed $\gamma$-ray spectrum extending to the highest observed energies among SNRs is unexpected for the low-speed shock of a middle-aged system. The current maximum energy of accelerated protons can be estimated as $160v_{s,8}^2(B/10~\mu{\rm G})(t_{\rm age}/10^5~{\rm yr})$~TeV \cite{Yamazaki:2006uf}, where $v_{s,8}$ is the shock velocity in units of $10^8$ cm~s$^{-1}$, $B$ is the magnetic field strength, and $t_{\rm age}$ is the SNR age. The Sedov expansion of the SNR shock yields a current shock velocity of $\sim300$~km~s$^{-1}$ for W51C, further reproducing the radius of W51C under the same evolutionary assumption \cite{Leahy:2017nrs}. The average $B$ of W51C should be of the order of $\sim50$~$\mu$G to account for the non-thermal radio emission of the whole SNR shell \cite{MAGIC:2012anb}. Taking $v_{s,8}=0.3$, $B=50$~$\mu$G, and $t_{\rm age}=30$~kyr, the current maximum proton energy is only $\approx22$~TeV, insufficient to account for LHAASO observation. This implies that the direct interaction model is disfavored by the LHAASO measurements.

An alternative scenario is that the highest-energy $\gamma$ rays observed in W51C are generated by PeV CRs that escaped from the SNR shock at early stages and are now encountering the MC associated with W51B. Depending on the diffusion conditions occurring near the SNR shock, a significant number of high-energy CRs could still be distributed nearby and capable of illuminating the MCs at a sufficient level for detection \cite{Li:2011qx}. The escaped CRs themselves may induce magnetic turbulence and suppress the diffusion coefficient through streaming instability \cite{Malkov:2012qd}. In this context, this remnant was previously proposed as a potential PeVatron candidate \cite{2019gnsg.book.....C}: the steepness of the proton slope could then be reasonably interpreted by the spectral softening effect owing to the energy-dependent diffusion of protons \cite{celli2019}. In terms of conversion efficiency from bulk motion into non-thermal particles, the combination between the estimated explosion energy of W51C and the aforementioned proton energetics requires a $\approx3.6\%$ level, when the default $d$ and $n_H$ are used, in agreement with the standards SNR paradigm for the origin of Galactic CRs, that in turn needs for a 10\% efficiency. If $n_H$ were substantially higher, a smaller conversion efficiency would be expected.

Electrons are accelerated alongside protons at the shock front of W51C and are also able to generate $\gamma$ rays through bremsstrahlung radiation and IC scattering of background photons. However, the detection of the $\pi^0$-decay bump suggests that the GeV emission from W51C primarily arises from hadronic interactions \cite{Jogler:2016lav}. Furthermore, the maximum energy attainable by electrons is severely restricted by the synchrotron energy loss, which can be estimated by $14v_{s,8}(B/10~\mu{\rm G})^{-1/2}$~TeV \cite{Yamazaki:2006uf}. Taking the current shock velocity of $v_{s,8}=0.3$ and $B=50$~$\mu$G, a cutoff energy of only $\approx2$~TeV is derived for electrons. Although electrons may be accelerated to higher energies during the early evolutionary stages of the SNR, the synchrotron energy loss prevents them from surviving to the present day \cite{2021MNRAS.508.6142M}. E.g., electrons with energy of $100$~TeV have a lifetime of only $\approx30$~years in a magnetic field of $50$~$\mu$G. Therefore, leptonic radiation from the SNR is unlikely to provide a significant contribution to the high-energy $\gamma$-ray spectrum observed by LHAASO.

\subsection{The young star clusters}

Besides the SNR W51C, it is not possible to exclude a contribution to the $\gamma$ radiation from other powerful accelerators., e.g., the nearby massive star clusters. These systems could enable particle acceleration either at the collective Wind Termination Shocks (WTSs) developed in young compact systems or at the multiple supernova shocks occurring there. The W51 region hosts at least four young embedded stellar clusters observed as bright radio-continuum and far-infrared sources, namely G48.9-0.3, G49.2-0.3, G49.4-0.3, and G49.5-0.4 \cite{2004MNRAS.353.1025K}. All are young enough that their evolution is expected to be powered mainly by strong WTSs \cite{2022MNRAS.512.1275V}. In fact, within the first 3~Myr of their lifetime, the geometry and energetics of a compact star cluster are dominated by the action of the collective WTS. This is the case of all the YSCs observed in the W51 complex, for which we hence proceed to the determination of the system physical parameters, including the hadron acceleration efficiency, by following the models of \cite{1977ApJ...218..377W} and \cite{morlino2021}, respectively. In particular, starting from the mass values obtained in \cite{2004MNRAS.353.1025K}, we apply the approach defined by \cite{celli2023} for the computation of wind speed and kinetic energy of each cluster. We report in Tab.A.2 expected values for each system, including WTS radius $R_{\rm TS}$ and bubble size $R_{\rm b}$, both computed with a circumstellar density of 10 protons per cc, wind mechanical power $L_{\rm w}$ and proton maximum energy $E_{\rm max}$, in the assumption that 10\% of efficiency is achieved in converting wind pressure into magnetic energy density and MHD-like turbulence (i.e. Kraichnan, a scenario in between the most effective Bohm and the Galactic observed Kolmogorov). We also provide sky maps in Appendix A together with the observed LHAASO emission.  

Interestingly, we find that the star clusters G48.9-0.3 and G49.2-0.3 are alternative viable accelerators to reproduce observations in the UHE domain, both morphologically and in terms of maximum energy. We further compute the total energetics so far injected by the collective winds of these two systems, accounting for the mass-dependent lifetime of main sequence stars \citep{buzzoni2002}, obtaining respectively values of $E_{\rm w} \simeq 6.1 \times 10^{51}$~erg and $E_{\rm w} \simeq 4.8 \times 10^{51}$~erg for G48.9-0.3 and G49.2-0.3, implying reasonable conversion efficiencies into protons at the level of 2-3\% to explain the observed radiation in hadronic scenarios. 

\subsection{The nearby PWN}
\label{subsec:pwn}
The presence of a putative nearby PWN radiating $\gamma$ rays through IC scattering was previously discussed in \cite{MAGIC:2012anb}. This PWN candidate, discovered by X-ray observations, is located southwest of the interaction region between W51C and W51B and might be the leftover of the supernova explosion associated with W51C.
Its $\gamma$-ray emission has even been tentatively identified by MAGIC, mostly in sub-TeV data, as an additional source, distinct from the dominant emission arising at the shock-cloud interaction region \cite{MAGIC:2012anb}, despite the significance of this observation is not enough for a conclusive assertion. Such a feature is, however, absent in current LHAASO data. 

Further investigations were conducted with LHAASO data to determine whether the $\gamma$-ray morphology exhibits variations across the three energy bins presented in Fig.~\ref{fig:skymap}, which would serve to evaluate leptonic scenarios for the origin of the radiation. We find that, for energies above $100$~TeV, the centroid of the emission is located at $({\rm R.A.,Dec.})_{100\rm{TeV}}=(290.80^\circ\pm0.06^\circ_{\rm stat},14.06^\circ\pm0.07^\circ_{\rm stat})$, closer to the PWN candidate; however, this shift is less than a $2\sigma$ deviation from the overall fit result, and it is not statistically significant. Additionally, the source extent at the highest energy bin has been measured to amount to $\sigma_{\rm ext} = 0.07^\circ\pm0.10^{\circ}_{\rm stat}$. However, due to the measurement uncertainty, it is not possible to conclusively determine whether there is a significant shrinking in size with increasing energy. These arguments, together with discussions that we further elaborate upon in Appendix B, suggest that the $\gamma$-ray contribution from the PWN candidate to the LHAASO observed radiation is marginal.

\section{Conclusions}
\label{sec:conclu}
With over two years of data collection and highly sensitive observations, LHAASO has significantly detected $\gamma$-ray events covering 2 TeV up to 200 TeV emanating from the W51 complex. The observed spectrum, for the first time, reveals the presence of radiation extending beyond $100$~TeV with a significance above $5\sigma$ and exhibits a pronounced spectral bending at tens of TeV, deviating from a power-law form with a $>5\sigma$ significance. Morphological analyses indicate the centroid of the $\gamma$-ray source is close to the interaction region between the SNR W51C and the dense MCs in W51B, consistent with the lower-energy observations. Incorporating the Fermi-LAT spectral data, we find that the broadband $\gamma$-ray emission from $60$~MeV to $200$~TeV can be naturally interpreted by the interaction between the CRs accelerated by W51C and the MCs, using a straightforward pp-collision model. The notable spectral bending suggests an exponential cutoff at $\sim400$~TeV or a power-law break at $\sim200$~TeV in the CR proton spectrum, most likely providing the first evidence of SNRs serving as CR accelerators approaching the PeV regime. 

Regarding other potential accelerators within the W51 complex, we point out that $\gamma$ rays possibly emitted by the candidate PWN CXO J192318.5$+$140305 are unlikely to impact our conclusions. 
Two YSCs embedded within W51B (G48.9-0.3 and G49.2-0.3) may be theoretically viable to account for the most energetic $\gamma$ rays observed by LHAASO, while the relatively coarse angular resolution of LHAASO precludes precise source identification through morphological analysis. Instruments with a significantly improved angular resolution are required to definitively identify the main contributor of these photons; future imaging atmospheric Cherenkov telescopes, such as LACT \cite{Zhang:2023bsi}, ASTRI Mini-Array \cite{2019EPJWC.20901001S}, and CTAO\cite{CTA:2020hii}, are poised to play a crucial role in this domain. Moreover, should LHAASO be capable of conducting spectral measurements at even higher energies for the W51 complex in the future, it would also provide further valuable insights into the origin of the UHE photons.

\section*{Conflict of interest}

The authors declare that they have no conflict of interest.

\section*{Author Contributions}
Zhe Li is the convener of this study and leading the observation data analysis. K.Fang led the manuscript preparation, performed the hadronic model fitting, as well as SNR and PWN interpretation.  C. Liu analysed the WCDA data and cross-check. Silvia Celli discussed the young star cluster scenario. Felix Aharonian gave crucial contributions for the theoretical interpretation. Yinxue Chai (supervised by Shaoru Zhang and Zhe Li) initially analyzed the data part from KM2A. Hui Zhu provided multi-wavelength data on the W51 region. Songzhan Chen, Shaoqiang Xi, and Shicong Hu gave useful suggestions on the background and diffusion analysis procedure. Yang Su helped providing the MWISP data covered a region of $2^\circ \times 2^\circ$ centered at SNR W51C. The whole LHAASO collaboration contributed to the publication, with involvement at various stages ranging from the design, construction, and operation of the instrument, to the development and maintenance of all software for data calibration, data reconstruction, and data analysis. All authors reviewed, discussed, and commented on the present results and the manuscript. 

\section*{Acknowledgements}
The authors would like to thank all staff members working year-round at the LHAASO site above 4400 meters above sea level to maintain the detector operating smoothly. We appreciate all LHAASO data processing team members for achieving high-quality reconstructed data and air shower simulation data. We are grateful to the Chengdu Management Committee of Tianfu New Area for the constant financial support for research with LHAASO data. 

This work is supported by: 
National Natural Science Foundation of China No.12393851, No. 12261160362, No.12393852, No.12393853, No.12393854, No. 12022502, No. 12205314, No. 12105301, No. 12105292, No. 12105294, No. 12005246, No. 12173039, Department of Science and Technology of Sichuan Province, China No.24NSFJQ0060, No.2024NSFSC0449, Project for Young Scientists in Basic Research of Chinese Academy of Sciences No. YSBR-061, and in Thailand by the National Science and Technology Development Agency (NSTDA) and National Research Council of Thailand (NRCT): High-Potential Research Team Grant Program (N42A650868).

This research made use of the data from the Milky Way Imaging Scroll Painting (MWISP) project, which is a multiline survey in 12CO/13CO/C18O along the northern Galactic plane with the PMO-13.7 m telescope. We are grateful to all the members of the MWISP working group, particularly the staff members at the PMO-13.7 m telescope, for their long-term support. MWISP was sponsored by the National Key R\&D Program of China with grants, 2023YFA1608000 and 2017YFA0402701 and the CAS Key Research Program of Frontier Sciences with grant QYZDJ-SSW-SLH047.




\clearpage
\newpage

\section*{LHAASO Collaboration}
Zhen Cao$^{1,2,3}$,
F. Aharonian$^{4,5}$,
Axikegu$^{6}$,
Y.X. Bai$^{1,3}$,
Y.W. Bao$^{7}$,
D. Bastieri$^{8}$,
X.J. Bi$^{1,2,3}$,
Y.J. Bi$^{1,3}$,
W. Bian$^{9}$,
A.V. Bukevich$^{10}$
% ...
\\
$^{1}$ Key Laboratory of Particle Astrophysics \& Experimental Physics Division \& Computing Center, Institute of High Energy Physics, Chinese Academy of Sciences, Beijing 100049, China\\
$^{2}$ University of Chinese Academy of Sciences, Beijing 100049, China\\
$^{3}$ TIANFU Cosmic Ray Research Center, Chengdu,  China\\
$^{4}$ Dublin Institute for Advanced Studies, 31 Fitzwilliam Place, 2 Dublin, Ireland \\
$^{5}$ Max-Planck-Institut for Nuclear Physics, P.O. Box 103980, Heidelberg 69029, Germany\\
$^{6}$ School of Physical Science and Technology \&  School of Information Science and Technology, Southwest Jiaotong University, Chengdu 610031, China\\
$^{7}$ School of Astronomy and Space Science, Nanjing University, Nanjing 210023, China\\
$^{8}$ Center for Astrophysics, Guangzhou University,  Guangzhou 510006, China\\
$^{9}$ Tsung-Dao Lee Institute \& School of Physics and Astronomy, Shanghai Jiao Tong University,Shanghai 200240, China\\
$^{10}$ Institute for Nuclear Research of Russian Academy of Sciences, Moscow 117312, Russia\\
% ...

\clearpage
\newpage

\textbf{\large {Supplementary materials} }


\appendix

\section{The young star clusters}
\label{sec:ysc}
\begin{figure}[h!]
  \centering
   \begin{adjustbox}{width=1.2\textwidth,center}
    \begin{minipage}{0.6\textwidth}
    \includegraphics[width=\linewidth, trim={2cm 0cm 1cm 2cm}, clip]{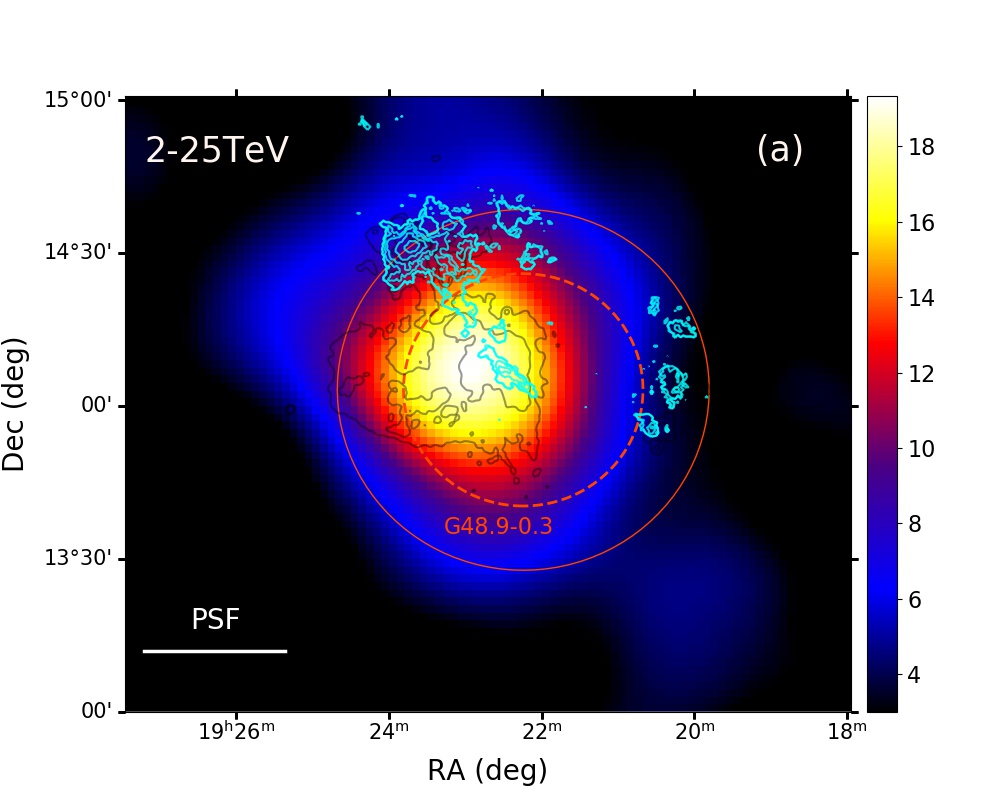}
    \end{minipage}%
    \begin{minipage}{0.6\textwidth}
    \includegraphics[width=\linewidth, trim={2cm 0cm 1cm 2cm}, clip]{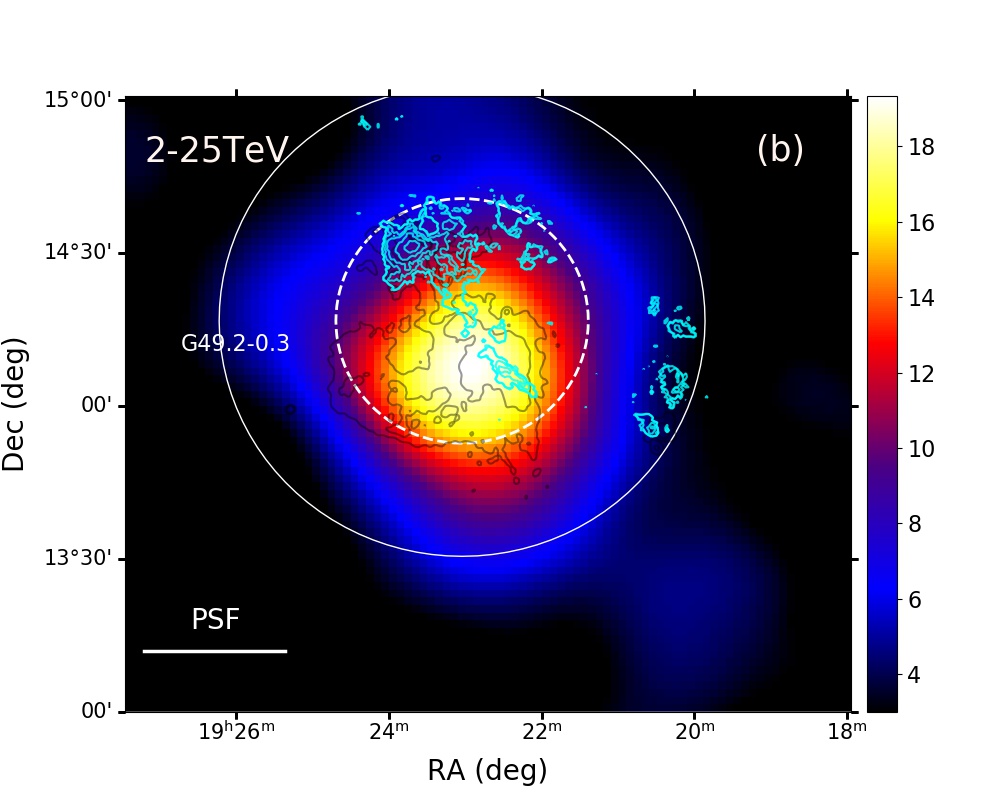}
    \end{minipage}
    \end{adjustbox}
    \begin{adjustbox}{width=1.2\textwidth,center}
    \begin{minipage}{0.6\textwidth}
    \includegraphics[width=\linewidth, trim={2cm 0cm 1cm 2cm}, clip]{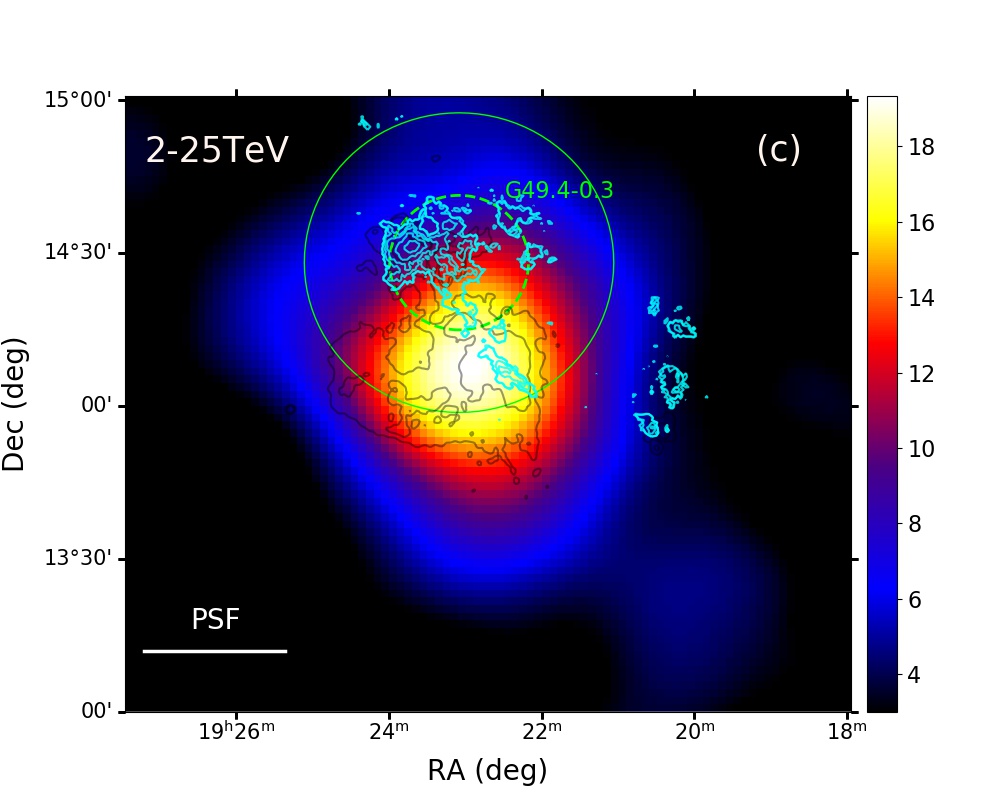}
    \end{minipage}%
    \begin{minipage}{0.6\textwidth}
    \includegraphics[width=\linewidth, trim={2cm 0cm 1cm 2cm}, clip]{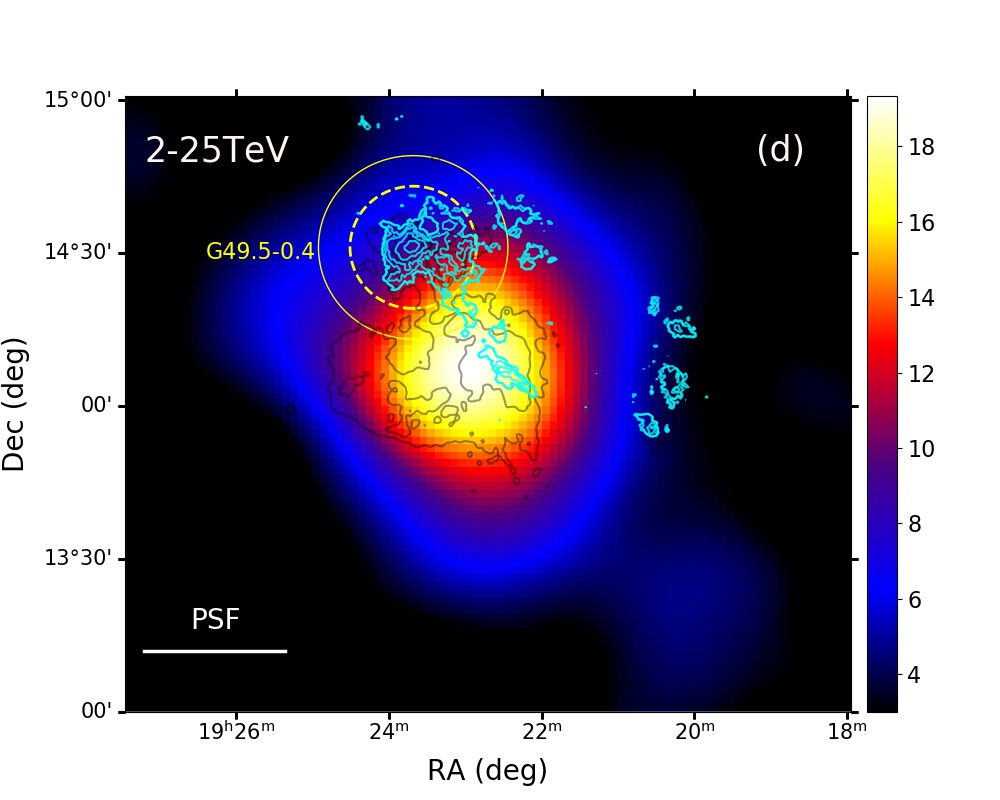}
    \end{minipage}
    \end{adjustbox}
    
\caption{Embedded YSCs in the W51 field, overlaid to LHAASO map in the $[2-25]$~TeV energy range: the dashed circle represents the WTS surrounding every cluster, while the solid circle represents the bubble radius. LHAASO PSF is also shown for comparison.(Color online)}
  \label{fig:2-25TeVclusters}
\end{figure}

\begin{figure}[h!]
  \centering
   \begin{adjustbox}{width=1.2\textwidth,center}
    \begin{minipage}{0.6\textwidth}
    \includegraphics[width=\linewidth, trim={2cm 0cm 1cm 2cm}, clip]{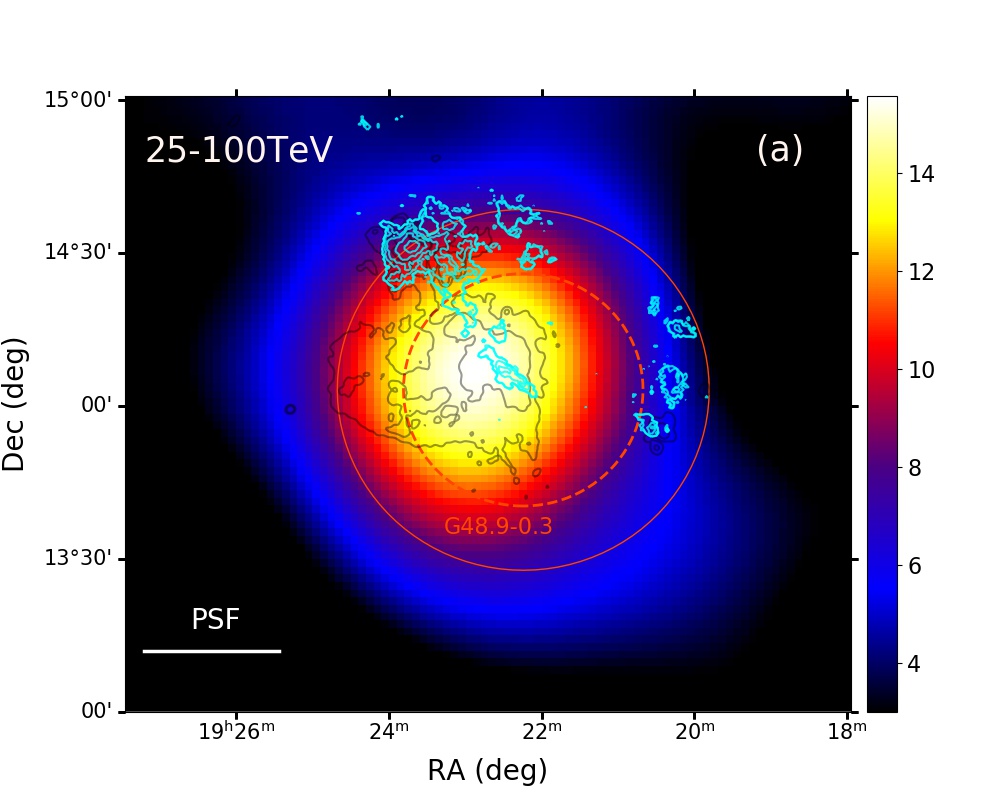}
    \end{minipage}%
    \begin{minipage}{0.6\textwidth}
    \includegraphics[width=\linewidth, trim={2cm 0cm 1cm 2cm}, clip]{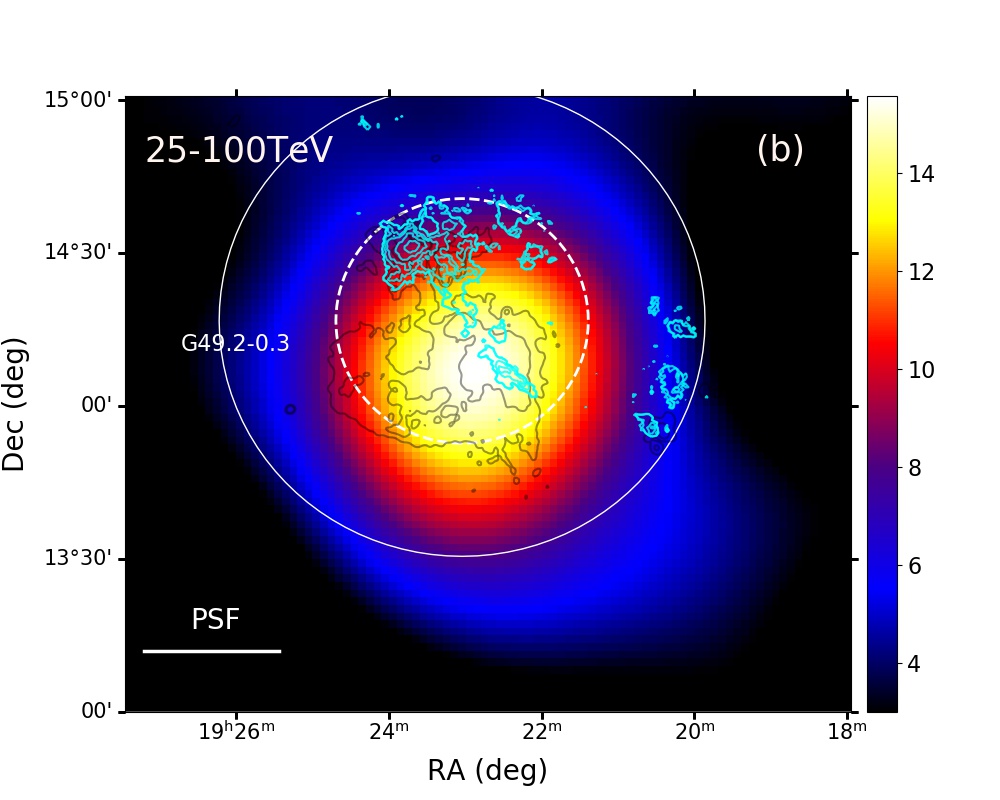}
    \end{minipage}
    \end{adjustbox}
    \begin{adjustbox}{width=1.2\textwidth,center}
    \begin{minipage}{0.6\textwidth}
    \includegraphics[width=\linewidth, trim={2cm 0cm 1cm 2cm}, clip]{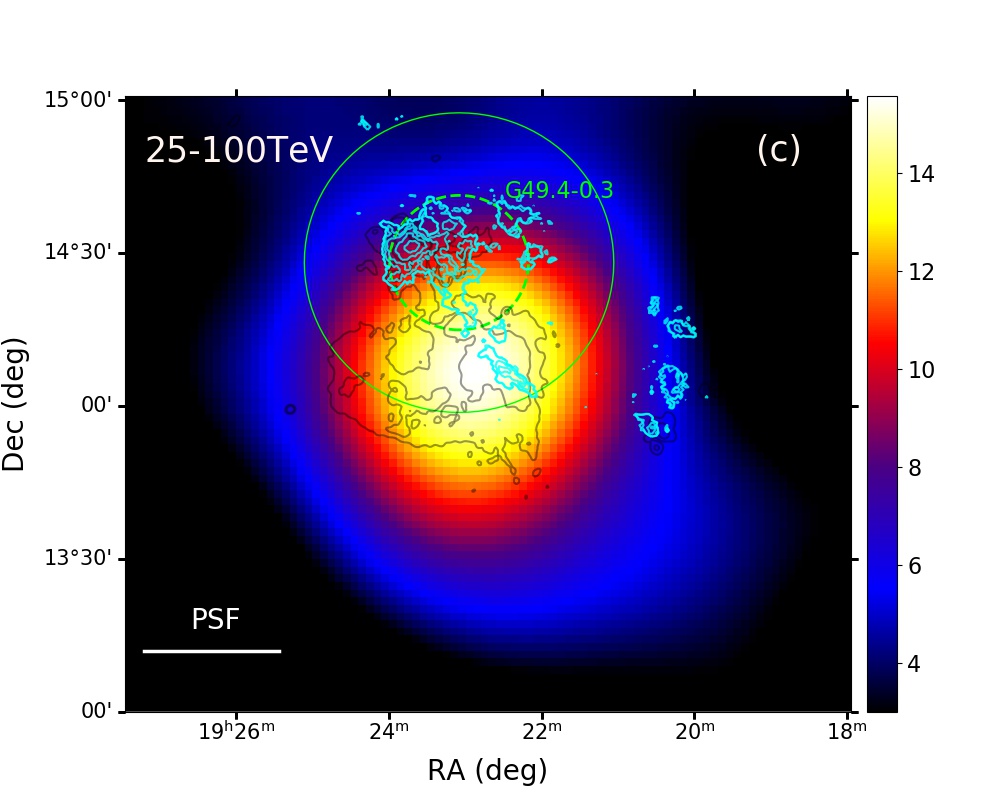}
    \end{minipage}%
    \begin{minipage}{0.6\textwidth}
    \includegraphics[width=\linewidth, trim={2cm 0cm 1cm 2cm}, clip]{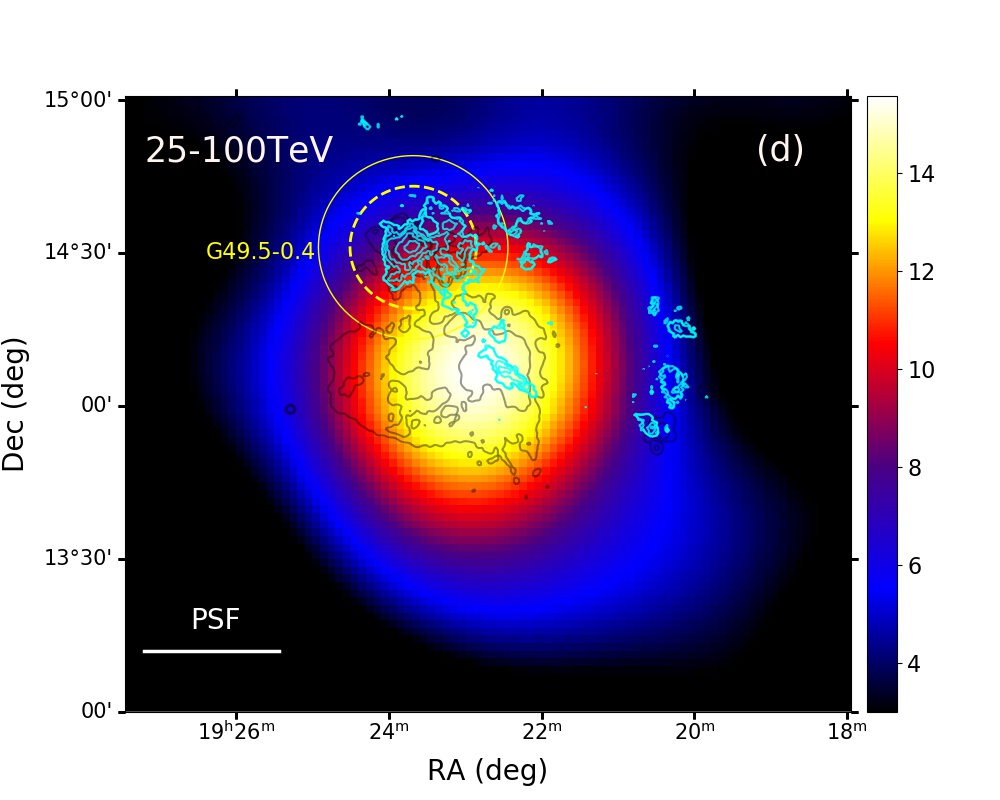}
    \end{minipage}
    \end{adjustbox}
    \caption{Same as Fig.~\ref{fig:2-25TeVclusters}, but in the $[25-100]$~TeV energy range.(Color online)}
  \label{fig:25-100TeVclusters}
\end{figure}

\begin{figure}[h!]
  \centering
    \begin{adjustbox}{width=1.2\textwidth,center}
    \begin{minipage}{0.6\textwidth}
    \includegraphics[width=\linewidth, trim={2cm 0cm 1cm 2cm}, clip]{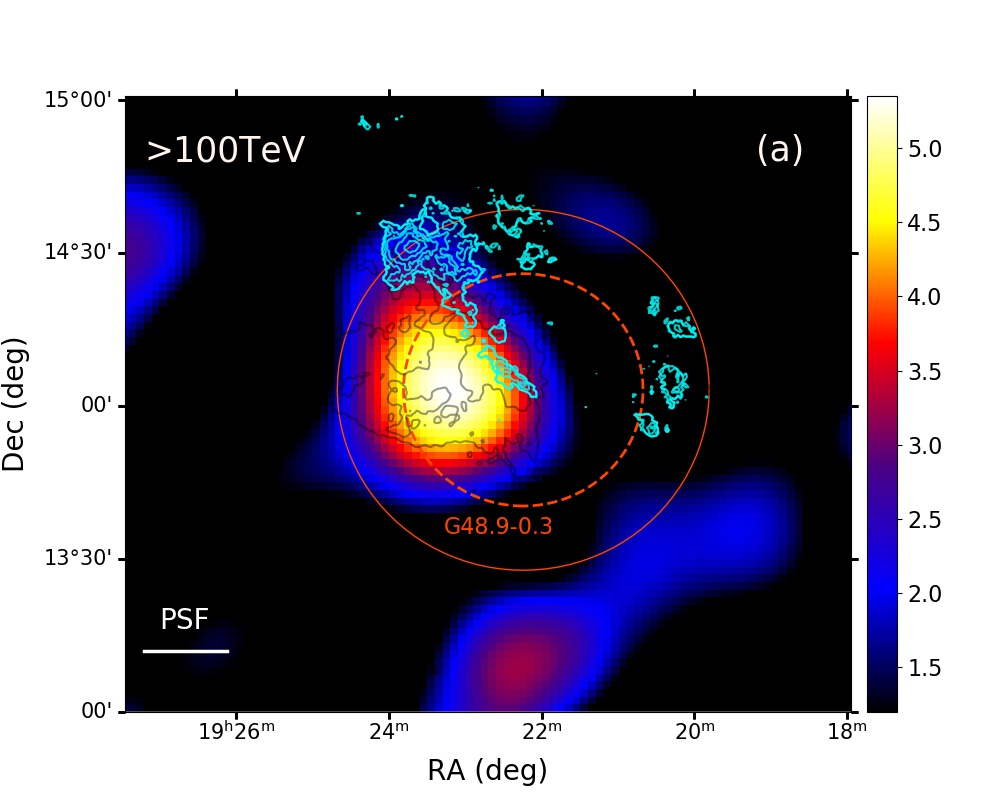}
    \end{minipage}%
    \begin{minipage}{0.6\textwidth}
    \includegraphics[width=\linewidth, trim={2cm 0cm 1cm 2cm}, clip]{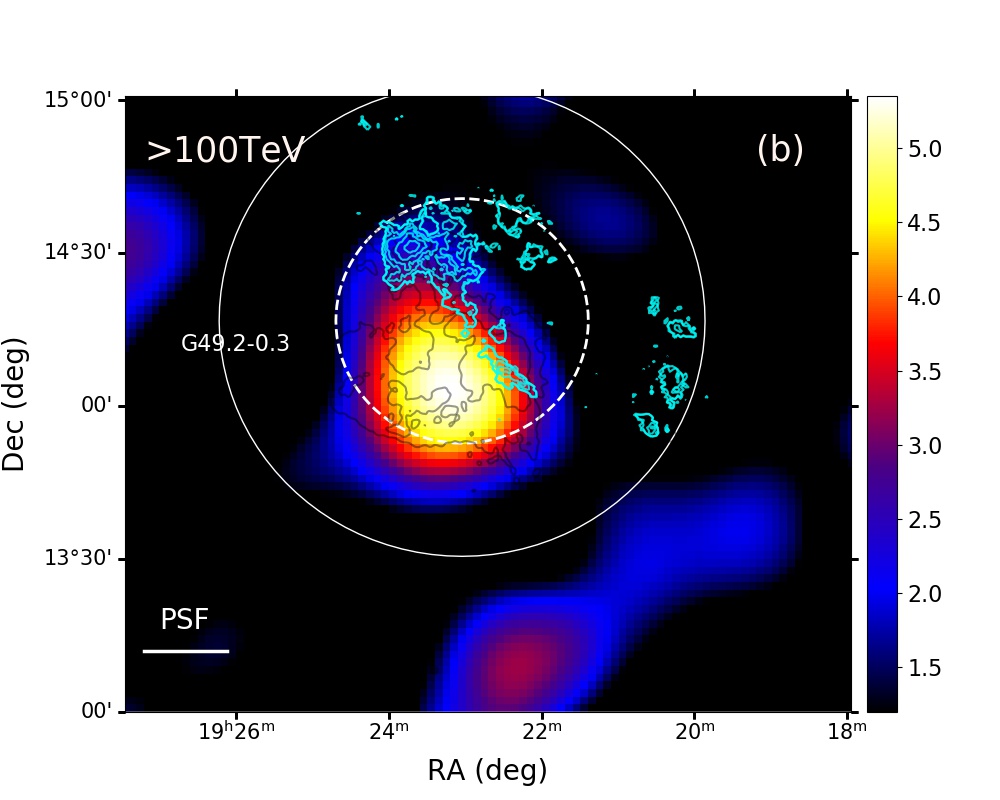}
    \end{minipage}
    \end{adjustbox}
    \begin{adjustbox}{width=1.2\textwidth,center}
    \begin{minipage}{0.6\textwidth}
    \includegraphics[width=\linewidth, trim={2cm 0cm 1cm 2cm}, clip]{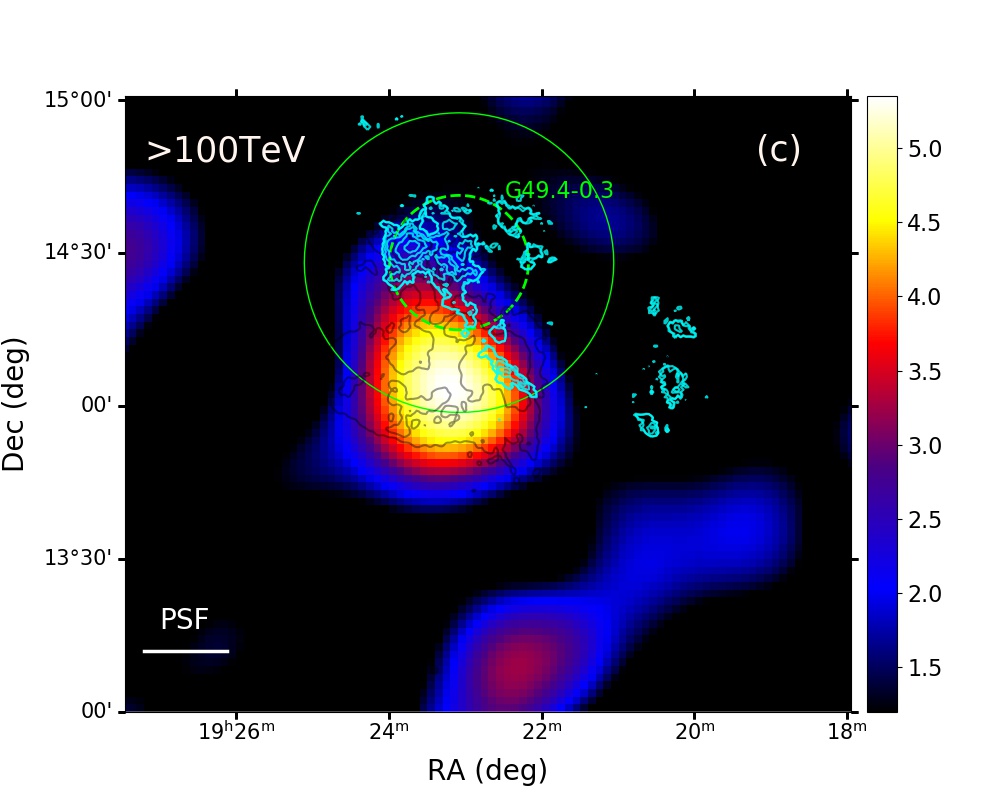}
    \end{minipage}%
    \begin{minipage}{0.6\textwidth}
    \includegraphics[width=\linewidth, trim={2cm 0cm 1cm 2cm}, clip]{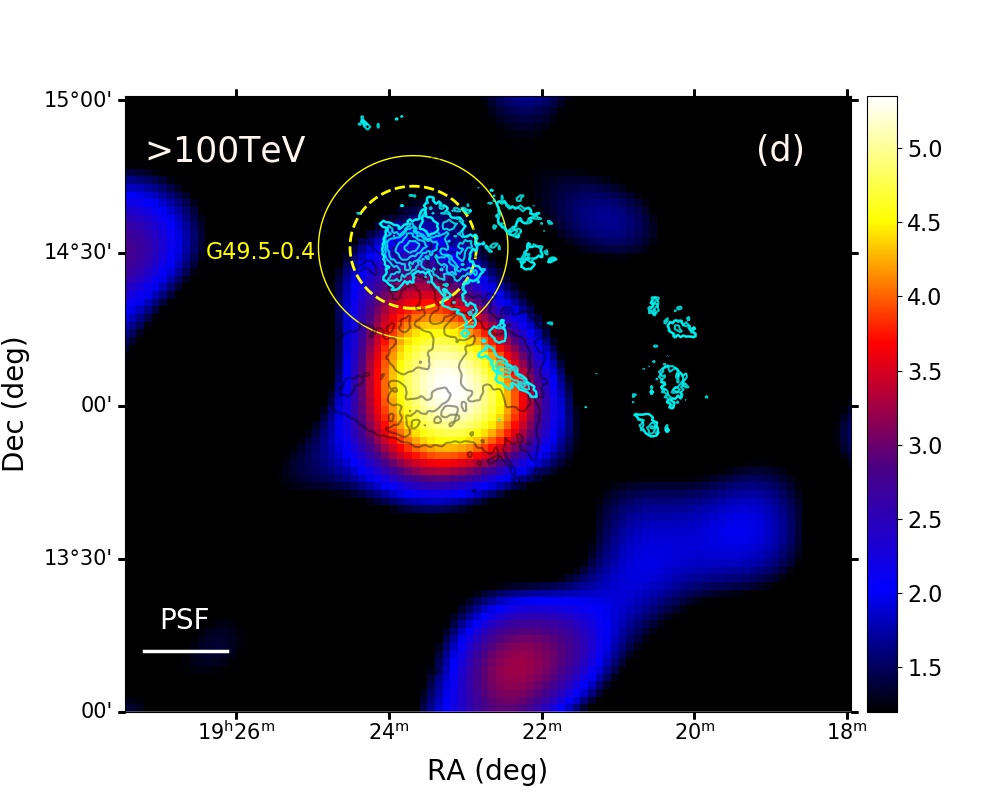}
    \end{minipage}
    \end{adjustbox}
  \caption{Same as Fig.~\ref{fig:2-25TeVclusters}, but in the $>100$~TeV energy range.(Color online)}
  \label{fig:100TeVclusters}
\end{figure}

Within the scenario of particle accleration at the WTS, two YSCs are found to potentially produce hadrons beyond LHAASO observations, namely G48.9-0.3 and G49.2-0.3. We  report in Figs.~\ref{fig:2-25TeVclusters}, \ref{fig:25-100TeVclusters}, and \ref{fig:100TeVclusters} the LHAASO excess map with respect to the cluster positions. 

\begin{table*}[h!]
   \centering
   \caption{Embedded star clusters observed in the W51 field: age $T_{\rm age}$ and mass $M_{\rm c}$ are taken from \cite{2004MNRAS.353.1025K}, while the following parameters are computed in the context of the WTS model, i.e. wind luminosity $L_{\rm w}$, WTS radius $R_{\rm TS}$, bubble radius $R_{\rm b}$, and expected proton maximum energy $E_{\rm max}$ in the Kraichnan diffusion scenario. All clusters were assumed in \cite{2004MNRAS.353.1025K} to be located at a distance of 6.5~kpc, in order to reproduce stellar data.}
   \begin{tabular}{lccccccc}
     \toprule
     YSC & $T_{\rm age}$ [Myr] & $M_{\rm c}$ [$M_\odot$] & $L_{\rm w}$ [erg/s] & $R_{\rm TS}$ [pc] & $R_{\rm b}$ [pc] & $E_{\rm max}$ [TeV] \\
     \hline
     G48.9-0.3 & 1.4 & $10000$ & $1.4 \times 10^{38}$ & $36.6$ & $56.9$ & 652 \\
     G49.2-0.3 & 3.0 & $8000$ & $5.1 \times 10^{37}$ & $38.0$ & $73.7$ & 377 \\
     G49.4-0.3 & 2.5 & $2000$ & $8.9 \times 10^{36}$ & $21.2$ & $46.7$ & 85.3 \\
     G49.5-0.4 & 0.7 & $4400$ & $3.9 \times 10^{37}$ & $19.3$ & $29.2$ & 173 \\
     \bottomrule
   \end{tabular}  
   \label{tab:sc}
 \end{table*}

\section{The putative pulsar wind nebula/pulsar halo}
\label{sec:pwn}
A solid proof of the existence of an associated PWN to the SNR W51C has not been provided yet: in fact, the corresponding pulsar is not observed, either because too faint or because its beam is not pointing toward us, implying that no estimate of their distances is available and hence no association can be claimed. \cite{2005ApJ...633..946K} have further attempted to evaluate the rotational power of the pulsar $\dot{E}$ starting from the observed X-ray luminosity $L_{\rm x} (0.5-10 \, {\rm keV}) = 5.9 \times 10^{34}$~erg~$s^{-1}$ of the surrounding nebula by means of an empirical relation derived for isolated pulsars \cite{1988ApJ...332..199S}, claiming a value as high as $\dot{E}=1.5 \times 10^{36}$~erg/s. We remark that such a value relies on the usage of a broadly scattered $L_x - \dot{E}$ correlation, derived from a limited sample of isolated pulsars, and it should hence be carefully regarded. The possibility that the highest energy radiation observed by LHAASO is partly contributed by this PWN, either from the nebula or from its pulsar halo, is explored here. 

The lack of a significant energy-dependent morphology in LHAASO data, which has to be expected in most leptonic-dominated scenarios, implies that diffusion should proceed in the Bohm domain, namely $D(E) \propto E$, such that the distance $R_{\rm d} \sim \sqrt{\tau_{\rm cool} (E) D(E)}$ propagated by electrons of energy $E$ in their synchrotron/IC cooling time $\tau_{\rm cool} \propto 1/E$ becomes energy independent. Moreover, X-ray observations of the compact object and surrounding nebula constrain the ambient magnetic field. E.g. Chandra observations in the $0.2-4$~keV band have in fact revealed a point-like source and a diffuse emission (the possible PWN) extending for about 5~pc, with flux as strong as $1.8 \times 10^{-12}$~erg~cm$^{-2}$~s$^{-1}$ \cite{2005ApJ...633..946K}. In order for the same X-ray emitting electrons/positrons (in the following simply addressed as electrons) to be also producing the flux of $\sim 200$~TeV $\gamma$ rays observed by LHAASO, i.e. $\sim 5 \times 10^{-14}$~erg~cm$^{-2}$~s$^{-1}$, a magnetic field at the level of $10 \, \mu$G is needed. 
Bohm diffusion in such a field corresponds to $D(100 \, {\rm TeV})=3\times 10^{26}$~cm$^2$/s: in order to reproduce the current radial extent of LHAASO observations of $\sim 20$~pc in terms of electrons diffusion length, a suppression of such diffusion coefficient is needed, amounting to a factor 35 at 100~TeV.
While even larger reductions with respect to the average Galactic diffusion have been claimed for other observed pulsar halos \cite{Abeysekara:2017old,Fang:2022fof}, the requirement, in this case, is more challenging because of the coupling between the maximum scattering rate and magnetic turbulence as strong as $350 \, \mu$G, at odd with the system age. Therefore, the pulsar halo scenario appears implausible.

With regards to the PWN scenario, the expected maximum energy of accelerated electrons can be constrained through the rotational power of the putative pulsar. In fact, by assuming the termination shock (TS) to be $R_{\rm TS} \simeq 0.1$~pc in size, the $B_{\rm TS} \simeq 10 \, \mu$G magnetic field translates into a constraint about the magnetic conversion efficiency at the TS equal to $\eta_{\rm B}=B_{\rm TS}^2R_{\rm TS}^2c/(2\dot{E}) \simeq 0.1$. This value determines the maximum energy of photons produced by accelerated electrons in IC scattering off the Cosmic Microwave Background (CMB), namely $E_{\gamma, \rm max}=0.9 \eta^{1.3}_{\rm e} \eta^{0.65}_{\rm B} \dot{E}^{0.65}_{36}$~PeV \cite{2022ApJ...930L...2D}, where $\eta_e$ is the conversion efficiency into electron/positron pairs. The LHAASO observed value of $E_{\gamma, \rm max}=200$~TeV hence requires $\eta_{\rm e} \simeq 0.8$, which is quite extreme with respect to standard PWNe. In other words, the expected maximum energy of leptons from the putative PWN is inconsistent with the highest energy observations of LHAASO, except for extreme values of the conversion efficiency  of the pulsar rotational power into electron/positron pairs. 

\begin{figure}[h!]
  \centering
  \includegraphics[width=0.65\textwidth]{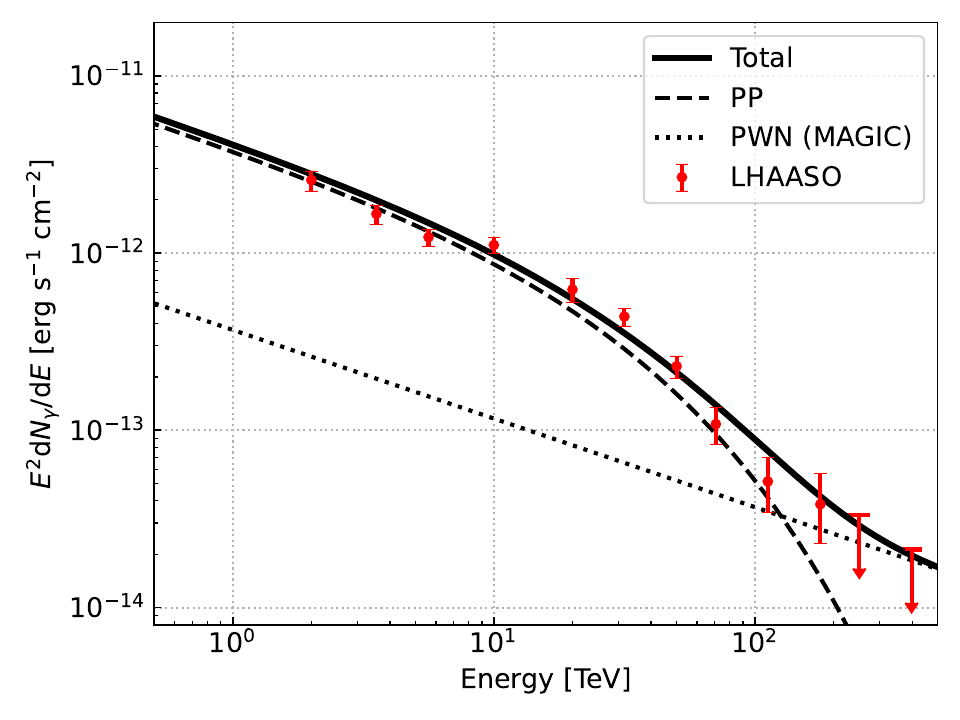}
  \caption{Hybrid model including the $\pi^0$-decay emission from the SNR-MC interaction and a possible gamma-ray PWN/pulsar halo revealed by MAGIC. The black solid line represents the maximum likelihood model obtained by fitting the LHAASO and Fermi-LAT data. We emphasize that this gamma-ray spectrum of the PWN/pulsar halo has been \textit{extrapolated} to $200$~TeV to assess the maximum potential influence of this source on the current study; the MAGIC measurement on this potential source is actually in the energy range of $0.35-2$~TeV.(Color online)}
  \label{fig:model_pwn}
\end{figure}

As addressed in Section~3.3, MAGIC has tentatively resolved a gamma-ray source at the site of the PWN candidate despite the relatively low significance level. The gamma-ray spectrum of this potential PWN, within the $0.35-2$~TeV range, can be described by a power law as $2.3\times10^{-13}(E/{\rm 1~TeV})^{-2.5}$~cm$^{-2}$~s$^{-1}$~TeV$^{-1}$, accounting for $\approx1/5$ of the total spectrum \cite{MAGIC:2012anb}. However, the configuration of its higher-energy spectrum remains unknown. Whether this source is a PWN or a pulsar halo, we examine the maximum potential influence this component may have on the parameter estimation of the proton spectrum by extrapolating its spectrum to higher energies following the power-law form measured by MAGIC. Fig.~\ref{fig:model_pwn} illustrates the comparison between the best-fit model and the data. At the highest energy bin measured by LHAASO, the PWN/pulsar halo contribution exceeds that of the hadronic component. This suggests that in this hybrid model, the cutoff energy of the proton spectrum would be marginally lower than in a purely hadronic model. Nonetheless, the deduced cutoff energy for the incident proton spectrum stands at $E_{p,\rm cut}=272_{-42}^{+46}$~TeV, still achieving a $4\sigma$ significance above $100$~TeV.


\end{document}